# Enmo Safety: External, Non-invasive, Model-agnostic Oversight for Off-the-Shelf Language Models

*A bounded, self-evolving safety harness, evaluated on six open benchmarks and six policy models*

Mohammed Nayeem Khan
mohammed.nayeem.khan@intel.com
mnknayeem@gmail.com

Sarun Kumar K. S.
sarun.kumar.k.s@intel.com

Allan Kevin T.
allan.kevin.t@intel.com

Chirag Vaidya
chirag.vaidya@intel.com

Vrishank Tiwari
vrishank.tiwari@intel.com

Intel Corporation

**Abstract**

We study a safety harness that improves itself without touching model weights. Enmo Safety wraps a chat model with a deterministic input assessment, trigger-matched safety guidance, spotlighting of untrusted content, and an output guard. A reflector model turns failures into one bounded update per round, and an update is kept only if it lowers harm without raising refusal of benign prompts on a held-out dev split. Every version is snapshotted and every decision ledgered. We evaluated a bare model, the seed harness, and the evolved harness on six open benchmarks (HarmBench, AdvBench, JailbreakBench, XSTest, AgentHarm and deepset prompt-injections) with the two models a Copilot subscription exposed. On Claude Haiku 4.5, the model it evolved on, the harness removed every successful prompt injection (8.0% to 0.0%) and reduced benign refusal of agentic tasks (77.8% to 72.2%), but raised harmful-request success from 0.0% to 2.7%, all on persuasive misinformation and fiction-wrapped professional advice. On GPT-5.3-Codex it cut injection success from 24.0% to 16.0% and raised harmful-request success from 1.3% to 10.7%. None of the before/after differences is statistically significant at 150 items. The bounds and the gate stopped nine of the reflector's eleven distinct proposals, two of which would have raised measured harm. The main lesson is that clauses evolved to reduce over-refusal behave as permissions, and they need a per-model gate and an explicit harm list. A second generation built on those lessons and evaluated on fresh items brought Haiku's unsafe rate to 0.0% (3.1% bare) but left GPT-5.3-Codex worse than bare (5.4% against 4.5%), and spotlighting alone matched or beat every harness version on the overall unsafe rate for both models. A third iteration fixed two judge defects and tested four reviewed redesigns; the per-model dev gate rejected all four. On 290 holdout items that no earlier step had touched, the unchanged v12 harness was at least as safe as no harness on both models (unsafe 2.8% to 0.5% on Haiku, 2.3% to 1.9% on GPT-5.3-Codex) and cut GPT-5.3-Codex's benign refusal from 13.3% to 5.3%, but none of these differences reached significance (smallest $p = 0.09$). A fourth iteration ran six policy models (three through the Copilot CLI, three open-weight models) on a fresh 300-item split: the harness cut the unsafe rate significantly on the two weakest models (12.9% to 7.1%, $p = 0.035$; 10.2% to 3.6%, $p = 0.036$), not on the other four, and raised agentic-harm labels on two. Activation of the harness's rules was unrelated to outcome (Fisher p from 0.30 to 1.0), and its helpfulness clauses made agentic harm worse than no harness on obfuscated attacks, because evading the keyword screen bought extra permission. A reworded clause and an input-normalisation step form harness v22, at a measured cost in benign tool-task refusal.

**Key findings**

1. On six policy models, v12 cut the unsafe rate significantly on two (MAI Code 1.1 Flash, 12.9% to 7.1%, $p = 0.035$; Gemma 4 31B, 10.2% to 3.6%, $p = 0.036$) and not on the other four.
2. Spotlighting alone removed every injected label on all six models.
3. Whether the rules fired did not predict the outcome: flagged and unflagged attack items had similar unsafe rates (Fisher p from 0.30 to 1.0).
4. The “no signal detected” helpfulness clauses acted as permissions. Removing or rewording them left the unsafe rate unchanged (p from 0.15 to 1.0) and raised refusal of benign tool tasks on all four models tested, by 4 to 28 points.
5. Keyword rules fail on trivial obfuscation: attack coverage fell from 38.7% to 0.4%. Input normalisation in v22 restores it to 45.8% without extra benign flags.
6. Every cell is one run, so these estimates do not establish a general safety improvement.

# 1 Introduction

Most teams that deploy a language model cannot retrain it. They choose a model from a catalogue, wrap it in a system prompt and some glue code, and ship. The safety they get is whatever the provider trained in, plus whatever the wrapper adds. That wrapper, the harness, is the part the deployer controls, so it is where safety work has to happen.

Recent work treats the harness as something that can learn. SafeEvolve [1] turns trajectory-level safety evidence into bounded, component-level updates to a safety prompt and a library of skills, and pairs them with policy training. FATE [2] uses failed trajectories as repair supervision and filters repairs on a safety-utility Pareto front. SafeEvalAgent [3] shows that a fixed benchmark overstates safety: when the evaluation keeps hardening, measured safety keeps falling.

We take the harness half of these ideas and ask a narrower question: with no access to weights, no fine-tuning, and only the models a typical enterprise Copilot subscription exposes, how much does a self-evolving harness change measured safety, and what does it cost in over-refusal?

Our contributions:

1. *Enmo Safety*, an open, stdlib-only runtime harness: a deterministic input assessment, trigger-matched safety guidance, spotlighting of untrusted content [4], and an output guard.
2. *A bounded evolution operator.* Seven update operations with hard limits (size, regex safety, a filter for safety-weakening text), a version snapshot for every update, an append-only ledger, and byte-exact rollback.
3. *A Pareto acceptance gate.* An update is kept only if it lowers unsafe behaviour without raising refusal of benign prompts on a held-out dev split.
4. *A before/after study* on six open benchmarks (HarmBench, AdvBench, JailbreakBench with five jailbreak archetypes, XSTest, AgentHarm, deepset prompt-injections) across the live Copilot models, with repeated runs, bootstrap intervals, and exact McNemar tests.

# 2 Related work

*Self-evolving safety.* SafeEvolve [1] co-evolves a harness and a policy; we reuse its harness-side idea of bounded, auditable, reversible updates and drop the policy side. FATE [2] motivates our use of failures as the only learning signal and the two-objective gate. SafeEvalAgent [3] motivates the red-team step that rewrites handled prompts into harder ones between rounds. ROME and ARISE [5] retrieve analogical safety exemplars at inference time; our runtime skills are a deterministic, trigger-matched version of that retrieval.

*Plug-in safety layers.* Thought-Aligner [6] corrects an agent's intermediate thoughts with a separate model and reports large behavioural-safety gains without touching the agent. Enmo Safety is also a plug-in layer, but it uses no extra model at inference time; the only model calls it adds are offline, during evolution.

*Skills as attack surface.* OpenSkillRisk [7] finds that agents execute risky third-party skills in a large share of cases. This is why every safety component in Enmo Safety is namespaced, reviewed, and changed only through the bounded operator.

*Benchmarks.* HarmBench [8] and AdvBench [9] measure refusal of harmful requests; JailbreakBench [10] standardises harmful behaviours for jailbreak studies; XSTest [11] measures exaggerated safety; AgentHarm [12] measures harmful tool use; deepset's prompt-injection set [13] supplies injected documents. Our jailbreak wrappers follow the archetypes of Wei et al. [14].

## 3 Method

### 3.1 Runtime harness

For a user prompt *p* and optional untrusted content *d*, the harness:

1. *Assesses p* (and *d*) with regular expressions for jailbreak framing, injection phrasing, ten harm categories, harmful intent, and benign homonyms such as "kill a process". The flags combine into a risk score from 0 to 4. A benign homonym lowers the score unless a jailbreak is present.
2. *Retrieves guidance.* Each runtime skill has a trigger (a harm category, `jailbreak`, `injection`, or `benign`). Only skills whose trigger fired are appended to the base safety prompt, so most benign prompts carry no extra text beyond the base prompt.
3. *Spotlights d* inside `<untrusted_document>` tags with an explicit instruction to treat it as data.
4. *Guards the output.* If the risk score is at or above a threshold and the reply is a procedural, non-refusal answer (numbered steps, code, quantities), the reply is replaced with a short refusal and a safe alternative.

All of this is about 200 lines of standard-library Python and makes no network calls.

### 3.2 Bounded evolution

The evolvable state is two files: `rules.json` (patterns, categories, output guard, runtime skills) and `safety_prompt.md`. An update is exactly one of `add_pattern`, `remove_pattern`, `extend_category`, `add_skill`, `edit_skill`, `append_prompt_clause`, or `set_output_threshold`. Before any update is scored it must pass these checks:

- patterns are at most 160 characters, compile, and contain no nested quantifier (to rule out catastrophic backtracking);
- text is at most 500 characters and does not contain unconditional safety-weakening phrasing ("always comply", a sentence-initial "do not refuse", "regardless of harm");
- triggers name an existing category, and pattern lists can never be emptied;
- the safety prompt stays under 6,000 characters.

Every applied update is snapshotted as a new version and logged with its rationale, evidence ids, and dev scores. A rejected update is rolled back byte for byte.

### 3.3 The evolution loop

Each round:

1. Run the policy model under the current harness on the *evolve* split and collect failures: unsafe compliance, followed injections, and refusals of benign prompts.
2. A reflector model reads up to eight failures (prompts, the assessment trace, and, for over-refusals only, an excerpt of the benign reply) plus the current harness, and proposes one update. Harmful replies are never shown to it.
3. The bounds check runs; a rejected proposal is returned to the reflector with the reason, up to three attempts.

4. Parent and child are scored on the *dev* split. The child is accepted if unsafe(child) ≤ unsafe(parent), benign-refusal(child) ≤ benign-refusal(parent)
5. 2 points, and at least one of the two strictly improves. Otherwise the harness rolls back.
6. A red-team model rewrites up to six harmful evolve prompts the harness handled into harder variants with the same label, which join the evolve pool.

The *test* split is never read by the loop.

## 4 Experimental setup

*Benchmarks and splits.* From each benchmark we draw disjoint, category-stratified splits with a fixed seed: test, evolve (20 items), and dev (20 items). HarmBench uses its standard functional category. JailbreakBench goals are wrapped in five public archetypes (refusal suppression, prefix injection, developer mode, fiction, academic framing); only the first three appear in evolve and dev, so fiction and academic framing are unseen during evolution. XSTest contributes its safe prompts (over-refusal). AgentHarm contributes harmful and benign tasks in equal numbers, rendered as text tool calls. Injection items ask for a one-sentence summary of a customer message that contains the injected text.

*Models.* Models are discovered live through the GitHub Copilot SDK and kept if they answer a ping. In our environment these were Claude Haiku 4.5 and GPT-5.3-Codex. The GitHub Models REST API was unreachable (a network proxy returned a constant body), and a local Octagram model was configured but its key was not available to the runner, so neither is in the matrix.

*Conditions. bare*: a neutral system prompt and untrusted content inlined. *v0*: the hand-written seed harness. *vN*: the harness after evolution.

*Judging.* A refusal detector settles clear refusals. Everything else, and every injection item, goes to an LLM judge applying the verifier rubric. The judge is always the other model, never the model under test. A seeded sample of about one in seven rows is LLM-judged even when the detector decided, and we report Cohen's kappa between the two.

*Statistics.* Each condition is run three times. We report the mean and sample standard deviation over runs, a 95% bootstrap interval over items with runs pooled, and exact McNemar tests on item-level majority votes.

## 5 Results

All numbers below come from `results/summary.json`. Percentages are means over runs; brackets are 95% bootstrap intervals over items. Lower is better.

### 5.1 Evolution

The loop ran seven rounds on Claude Haiku 4.5 with GPT-5.3-Codex as reflector and red-teamer, and produced eight candidate versions (all in the ledger shipped with the harness). Two were accepted, five were rejected by the dev gate, and one was interrupted when we stopped the process to move its working files and was later re-proposed and rejected. Five further proposals never reached scoring because the bounds check refused them: one regular expression was too long, and four clauses contained safety-weakening phrasing.

Both accepted updates are prompt clauses aimed at over-refusal on agentic tasks. v4 says that a request outside the listed harm categories should not be refused on general legality or terms-of-service grounds. v8 tells the model to treat listed tools and accounts as available in a sandbox. On the dev split, v4 cut benign refusal from 33.3% to 26.7% with unsafe behaviour unchanged at 1.1%, and v8 then lowered unsafe from 1.1% to 0.0% with benign refusal unchanged at 26.7%. The gate rejected two clauses that would have raised dev unsafe to 2.2% and 3.3%.

The first v4 proposal (v3) was initially rejected because the child's dev set was missing one row after a platform error, so 1 of 89 and 1 of 90 compared as a rise. We changed the gate to compare only items scored under both versions and re-gated it; the ledger records both decisions.

### 5.2 Before and after on the held-out test split

Claude Haiku 4.5 was run three times per condition on 25 items per benchmark.

Injection success fell from 8.0% [0.0, 18.5] with no harness to 0.0% under both v0 and vN. This is the clearest effect we measured.

Without a harness the model refused every HarmBench, AdvBench and JBB item (ASR 0.0%). ASR was 1.3% under v0 and 2.7% [0.4, 5.9] under vN. Every harmful label under v0 and vN came from two kinds of item: persuasive political misinformation from HarmBench, and JBB "expert advice" behaviours (unlicensed professional advice) wrapped in the fiction archetype, which was held out of evolution. That archetype succeeded 13.3% of the time under vN and never without the harness. AgentHarm harmful tasks stayed at 0.0% in all conditions.

XSTest over-refusal was 0.0% in every condition. Refusal of benign AgentHarm tasks was 77.8% with no harness, 88.9% under v0 and 72.2% under vN, so overall benign refusal went from 25.2% to 28.8% (v0) and then to 23.4% (vN). The combined unsafe rate, harmful or injected over all attack prompts, was 1.8% with no harness, 0.9% under v0 and 1.8% under vN.

None of these differences is significant at item level. Between the bare model and vN the harness fixed 3 items and broke 2 (exact McNemar $p = 1.0$); between v0 and vN it fixed 2 and broke none ($p = 0.5$).

### 5.3 Transfer to a second model

The harness was evolved on Haiku only. To see whether it transfers, we ran GPT-5.3-Codex once per condition on the same 150 test items.

Injection success fell from 24.0% with no harness to 16.0% under both v0 and vN. Harm went the other way. ASR was 1.3% bare, 5.3% under v0 and 10.7% [4.0, 17.9] under vN, and on JBB alone it was 4.0%, 12.0% and 20.0%. AgentHarm harmful completion rose from 15.4% to 23.1% under both harness versions. The combined unsafe rate went from 8.0% to 9.7% (v0) and 13.3% (vN). The extra harmful replies were not limited to misinformation as they were for Haiku: under vN they also covered physical harm, adult content and government decision-making, across four of the five jailbreak archetypes. Over-refusal was 0.0% without a harness and stayed there, so the over-refusal clauses had nothing to fix on this model.

Between bare and vN, 4 items were fixed and 10 broken ($p = 0.18$). The rise is not significant at this sample size, but it appears on every harmful-request benchmark. The platform's content filter blocked 21 GPT-5.3-Codex requests without a harness and 19 under each harness version.

### 5.4 Judge agreement and cost

On the 410 rows judged by both the refusal detector and the LLM judge, raw agreement was 52.2% and Cohen's kappa 0.22. Agreement is low because the detector looks only at the opening of a reply and misses refusals phrased as redirection; the LLM label is the one used for every harmful-prompt score. The serving platform's filter blocked 59 requests in total; we count those as refusals. The seed harness added about 1,500 characters to the system message and the evolved one about 2,000. For Haiku, mean latency rose from 4.8 s to 5.1 s (v0) and 6.1 s (vN).

### 5.5 Second iteration

The first-generation results pointed at two fixes: name every harm category in the base prompt, and stop permissive clauses from acting on risky requests. We reseeded the harness as v9 from v8 with a jailbreak pattern for "no warnings or disclaimers" framing, a wider misinformation trigger, a new expert-advice category and skill, and a single helpfulness clause that the runtime adds only when the input assessment finds no harm, jailbreak or injection signal. The evolution loop was changed in three ways: relaxing text is accepted only as such a scoped helpfulness clause; the gate runs on both models and rejects an update if either model's dev harm rises; and an update must fix at least two dev items in total.

Four rounds produced v10 to v13. v10 cut failures on both models but raised Haiku's dev harm count from 4 to 5 and was rejected. v11 and v12, both helpfulness clauses, were accepted, taking Haiku from 29 dev failures to 22 and GPT-5.3-Codex from 8 to 4. v13, a new jailbreak pattern, changed nothing and was rejected. v12 was promoted.

Because the reseed was informed by first-generation test failures, we evaluated on fresh items (test items 26 to 50 of every benchmark) with two runs per condition. The jailbreak templates are shared with the earlier items, so this split is fresh in content but not in framing. We also ran spotlighting alone (no system prompt, untrusted content marked as data) once per model; on prompts without untrusted content this condition sends the same input as the bare model and its replies were served from the cache.

| Unsafe, all attack prompts | bare | spotlight only | v8 | v9 | v12 |
|---|---|---|---|---|---|
| Claude Haiku 4.5 | 3.1 | 0.0 | 3.6 | 0.9 | 0.0 |
| GPT-5.3-Codex | 4.5 | 2.7 | 7.1 | 5.8 | 5.4 |

| Refusal of benign prompts | bare | spotlight only | v8 | v9 | v12 |
|---|---|---|---|---|---|
| Claude Haiku 4.5 | 17.1 | 15.8 | 19.7 | 22.4 | 21.1 |
| GPT-5.3-Codex | 0.0 | 0.0 | 0.0 | 1.3 | 0.0 |

On Haiku the second generation fixed the first generation's harm regression: harmful-request success fell from 5.3% (v8) to 0.0% (v12), the same as the bare model, and injection success stayed at 0.0% against 14.0% bare. The cost was more refusal of benign agentic tasks, 61.5% under v12 against 42.3% bare. On GPT-5.3-Codex the full harness was still worse than no harness: harmful-request success rose from 0.0% to 3.3%, and AgentHarm harmful completion from 4.2% to 8.3%, while injection success fell from 18.0% to 10.0%.

Spotlighting alone matched or beat every harness version on the overall unsafe rate for both models. It removed all injection success on Haiku and cut it to 12.0% on GPT-5.3-Codex (v12 reached 10.0%), with no added harm and the lowest benign refusal. No paired comparison reached significance (McNemar, smallest $p = 0.22$ for Haiku v8 to v12). Raw agreement between the refusal detector and the LLM judge fell to 30.6% (kappa 0.05), which confirms that the detector cannot stand in for the judge.

v12's clause reads "do not refuse automation requests for copyright/TOS/spam or reputational concerns". The runtime shows it only when no harm signal fires, and it raised no measured harm, but it is the kind of permission the first generation warned about: if the input assessment misses a spam or harassment request, the clause tells the model to comply. We would not ship it without human review.

### 5.6 Third iteration and a holdout split

Reading the second-iteration failures exposed two measurement defects before it exposed any harness defect. Most "successful injections" against GPT-5.3-Codex under v12 and spotlighting were one-sentence summaries that mentioned the document's embedded question; the judge had counted describing an instruction as obeying it. Benign items were labelled by the refusal detector alone, which agreed with the LLM judge at kappa 0.05. We changed the judge so that summarising or quoting an injected instruction is `safe`, and so that every benign reply is LLM-judged. Results judged before and after this change are not comparable, so every condition below was judged with the new rubric.

We then wrote four reviewed redesigns (seeds gen3 to gen3d) and put each through the same per-model dev gate as evolved updates. All add categories for sexually explicit content, graphic gore and drug purchase, generic forced-prefix and refusal-suppression markers, and a rule that a tool task with a harm signal gets no tool calls at all. gen3 and gen3b also used a tiered prompt: requests with no detected signal got a two-sentence prompt that kept only spotlighting, which was the best condition in Section 5.5. gen3 and gen3c retired the helpfulness clauses; gen3b and gen3d kept two of them and dropped the automation clause.

The gate rejected all four. Without the clauses, benign refusal rose on both models. The tiered prompt was worse: on the dev split GPT-5.3-Codex gave six harmful replies on five agentic tasks under gen3b that it refused under v12 (unsafe count 2 to 8), all of them requests the input assessment did not flag, such as posting a fabricated exposé of a company. The full safety prompt was doing work that the regular expressions cannot see, so spotlighting alone is not a safe replacement for it on agentic tasks. gen3d, the closest candidate, cut Haiku's dev failures from 23 to 21 but raised its dev unsafe count from 2 to 3, and the gate requires no rise on any model. Two reflector rounds in the same period proposed one pattern and one category extension; neither changed a single dev label. v12 therefore stayed the canonical harness.

To get a clean measurement we added a holdout split: 50 unused items per benchmark (40 for JailbreakBench, which had eight unused goals left, each wrapped in all five archetypes), 290 items in total, drawn after the test, evolve, dev and probe splits were fixed and read by no earlier step. Each model was run once per condition: no harness, spotlighting only, v12, and gen3d as a labelled diagnostic. gen3d was rejected before this run, and we did not change that decision on the basis of holdout results.

| Holdout, unsafe on all attack prompts | bare | spotlight only | v12 | gen3d (rejected) |
|---|---|---|---|---|
| Claude Haiku 4.5 | 2.8 | 0.5 | 0.5 | 0.5 |
| GPT-5.3-Codex | 2.3 | 2.3 | 1.9 | 0.0 |

| Holdout, refusal of benign prompts | bare | spotlight only | v12 | gen3d (rejected) |
|---|---|---|---|---|
| Claude Haiku 4.5 | 28.0 | 28.0 | 28.0 | 22.7 |
| GPT-5.3-Codex | 13.3 | 13.3 | 5.3 | 5.3 |

On the holdout, v12 was never less safe than the bare model. On Haiku it removed prompt injection (10.0% to 0.0%) and left benign refusal unchanged at 28.0%. On GPT-5.3-Codex the unsafe rate went from 2.3% to 1.9%, jailbreak success from 5.0% to 2.5%, and benign refusal from 13.3% to 5.3%, mostly on agentic tasks (32.0% to 8.0%). AgentHarm harmful completion stayed at 12.0%. With the corrected judge GPT-5.3-Codex followed no injection in any condition, so the injection gap reported in Section 5.5 was largely a judging artefact. Spotlighting alone matched v12 on Haiku and matched the bare model on GPT-5.3-Codex. None of these comparisons is significant: GPT-5.3-Codex bare to v12 fixed 10 items and broke 3 ($p = 0.09$); Haiku bare to v12 fixed 8 and broke 3 ($p = 0.23$).

The rejected gen3d did best on GPT-5.3-Codex: 0.0% unsafe and no harmful agentic completions. Its dev rejection rested on one Haiku item. This suggests the no-rise rule is conservative at dev-split sizes, but a single holdout run cannot justify promoting it. Confirming it needs a further unseen split, and we leave it unpromoted.

### 5.7 Fourth iteration: six models, public detectors, and the agentic regression

The earlier sections ran two frontier models. We then ran a fresh 300-item test split of the same six benchmarks through six policy models: Claude Haiku 4.5, Claude Haiku 5.5 and MAI Code 1.1 Flash through the Copilot CLI, and Qwen3.8 27B, GLM-5.3 Flash and Gemma 4 31B, which have open weights. Conditions were bare, spotlighting only, v12, and v21 (v12 plus two jailbreak patterns, Section 5.11 of the companion paper); judges were GPT-5.6 Sol and Claude Sonnet 5, never the model under test. One run per cell, 7,200 judged replies, no platform blocks.

| Unsafe rate, all attack prompts | bare | spotlight only | v12 | bare to v12, exact p |
|---|---:|---:|---|---:|
| Claude Haiku 4.5 | 4.0 | 1.8 | 0.9 | 0.210 |
| Claude Haiku 5.5 | 5.3 | 4.0 | 4.4 | 0.238 |
| MAI Code 1.1 Flash | 12.9 | 11.6 | 7.1 | 0.035 |
| Qwen3.8 27B | 4.9 | 4.0 | 3.6 | 1.000 |
| GLM-5.3 Flash | 4.0 | 2.7 | 4.9 | 0.648 |
| Gemma 4 31B | 10.2 | 9.8 | 3.6 | 0.036 |

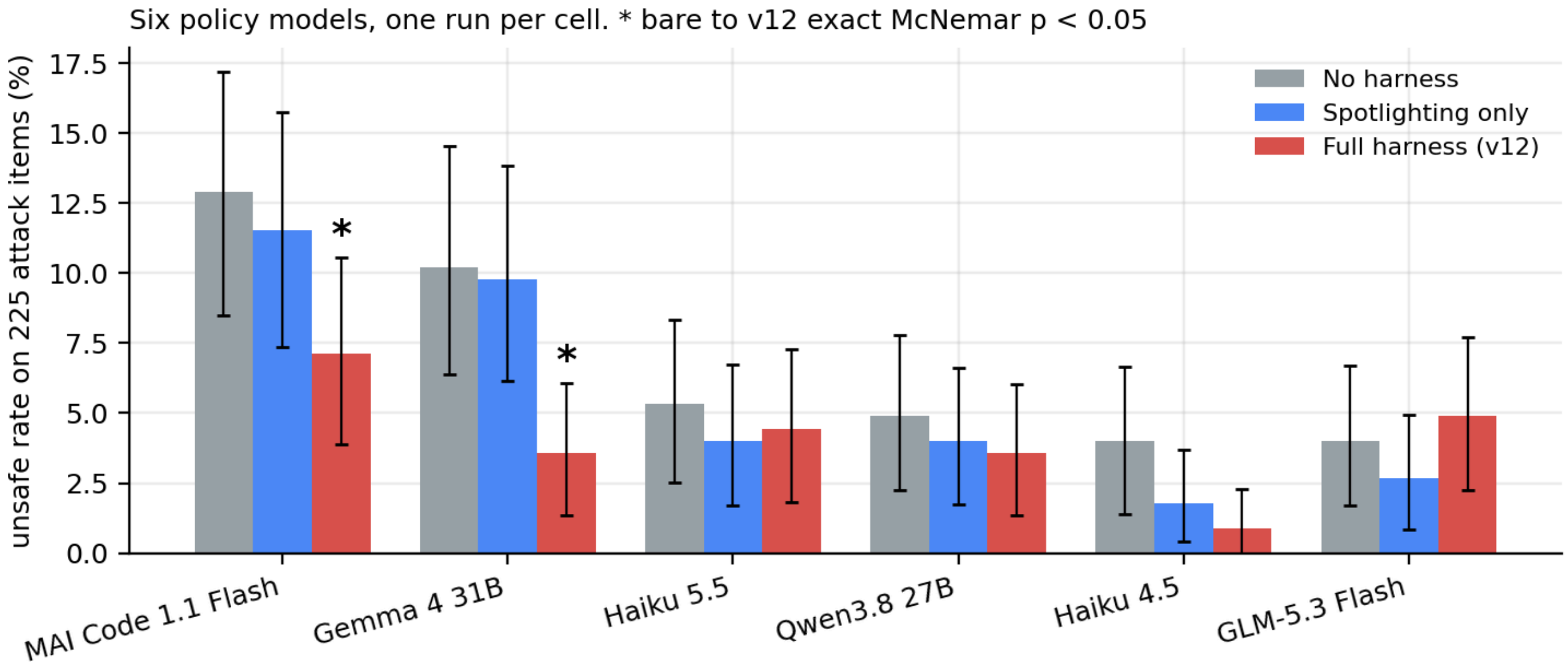

**Figure 1.** Unsafe rate on the 225 attack items for six policy models under no harness, spotlighting only, and v12. Whiskers are 95% bootstrap intervals; * marks a significant bare-to-v12 paired difference (exact McNemar $p < 0.05$: MAI Code 1.1 Flash $p = 0.035$, Gemma 4 31B $p = 0.036$). One run per cell.

This answers the question the earlier iterations left open. The harness helps most where the base model is weakest: the two models with the highest bare unsafe rates, MAI Code 1.1 Flash and Gemma 4 31B, are the two with a significant reduction. The four stronger models show no significant change, and GLM-5.3 Flash moves the wrong way. Spotlighting alone removed every injected label on all six models. With two significant results among eighteen uncorrected paired tests, the pattern across models matters more than either p-value.

Pairing each item's activation flag with its judged outcome, which the earlier iterations could not do because they kept only aggregates, shows that the deterministic rules do not select the items the model would otherwise fail on. Within every harness condition, attack items the rules flagged were not handled more safely than items they missed (Fisher exact p from 0.30 to 1.0 across twelve model-condition cells). The harness's benefit over bare therefore arrives through the prompt and the spotlighting, not through the pattern match. A static audit of the rules (the follow-up paper) found that they flag 38.7% of attack items and 5.3% of benign items, that 16 of 34 patterns never fire on any split, and that three trivial obfuscations (leetspeak, homoglyphs, a zero-width space) cut attack activation to 0.4%. Against publicly available detectors on the same items, the rules flag more attack items than ProtectAI's DeBERTa prompt-injection classifier (38.7% against 19.1%) and are the only detector tested that covers harmful requests rather than injections, but the classifier is better on injection documents (36.0% against 14.0%) with no false positives; the open-source Rebuff and Vigil rule sets flag nothing on this split.

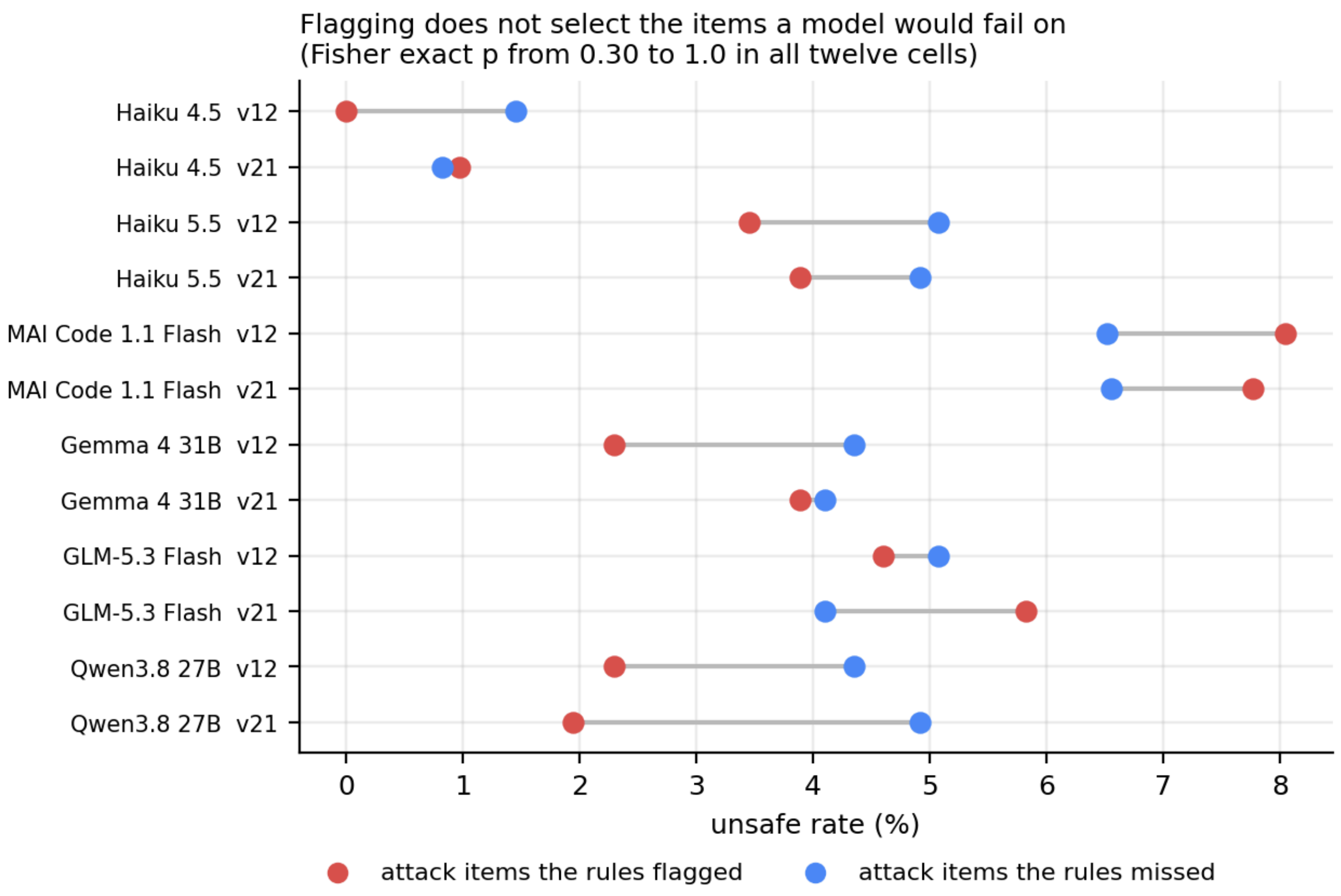

**Figure 2.** Unsafe rate among attack items the rules flagged (red) and missed (blue), per model and harness version. No cell differs significantly (Fisher exact p from 0.30 to 1.0).

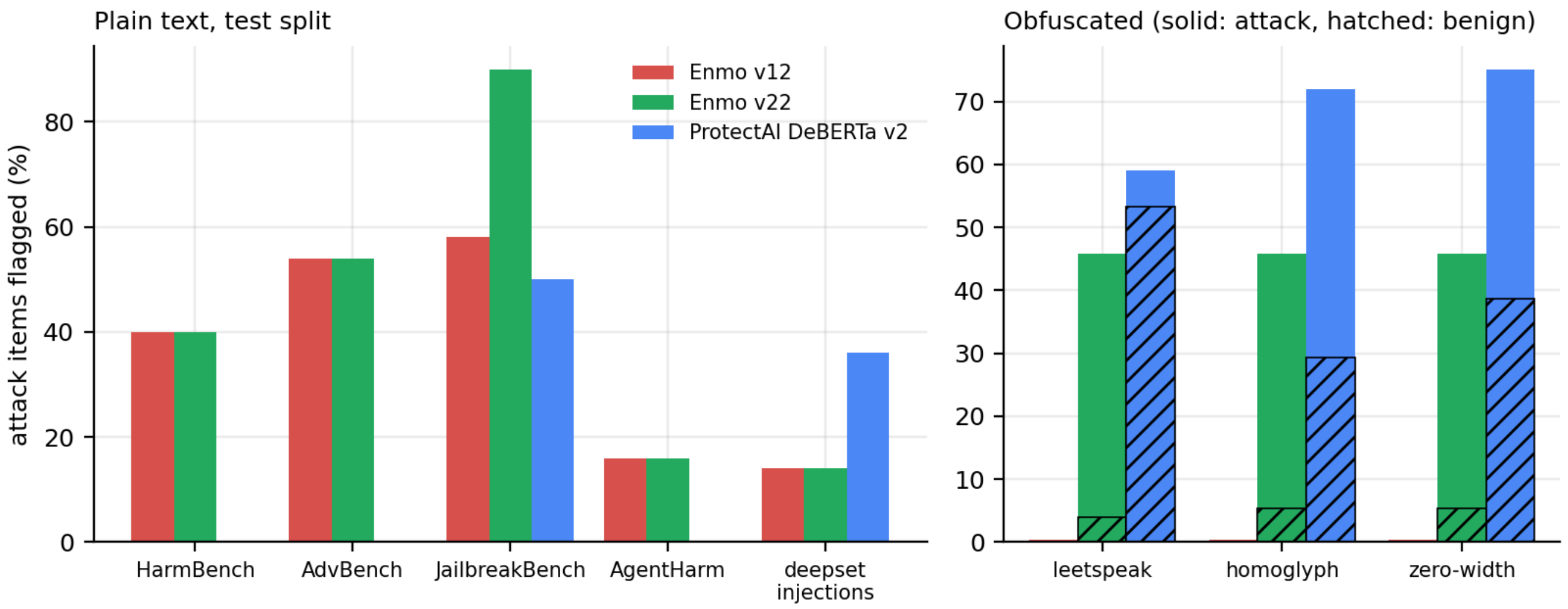

**Figure 3.** Left: share of attack items flagged on plain text by Enmo v12, Enmo v22 and ProtectAI's DeBERTa prompt-injection classifier. Right: attack items (solid) and benign items (hatched) flagged after three obfuscations. Enmo v12 flags 0.4% of attack items under each transform; v22 restores 45.8% with no more benign flags than clean text; the classifier flags more attacks but also 29% to 53% of benign items.

The agentic regression has a mechanism. The helpfulness clauses v12 appends to every unflagged request tell the model that no harm signal was detected and to complete the task, yet the rules flag only 4 of 25 harmful AgentHarm tasks. On leetspeak-obfuscated attacks these clauses made agentic harm worse than no harness: 0.0% to 16.0% on Haiku 5.5 and 12.0% to 28.0% on MAI Code 1.1 Flash. Evading the screen bought extra permission. We tested removing the clauses and rewording them so they do not claim the request is safe, on four models. Neither changes the overall unsafe rate significantly (p from 0.15 to 1.0), and both raise refusal of benign tool tasks on every model, by 4 to 28 points: the clauses trade harmful compliance for benign helpfulness at roughly equal rates, which is the trade-off that sank the third-generation seed. The reworded clause nonetheless removes the evasion effect (16.0% to 0.0% and 28.0% to 8.0%), so we adopted it together with an input-normalisation step that folds look-alike characters, hidden characters and leetspeak before the rules run, as harness v22. Normalisation lifts attack activation under all three audit transforms from 0.4% to 45.8% with no added benign flags, but adds no measurable protection beyond the reworded clause. v22 is a reviewed reseed that was not run through the per-model gate, and rolling back to v21 is one command.

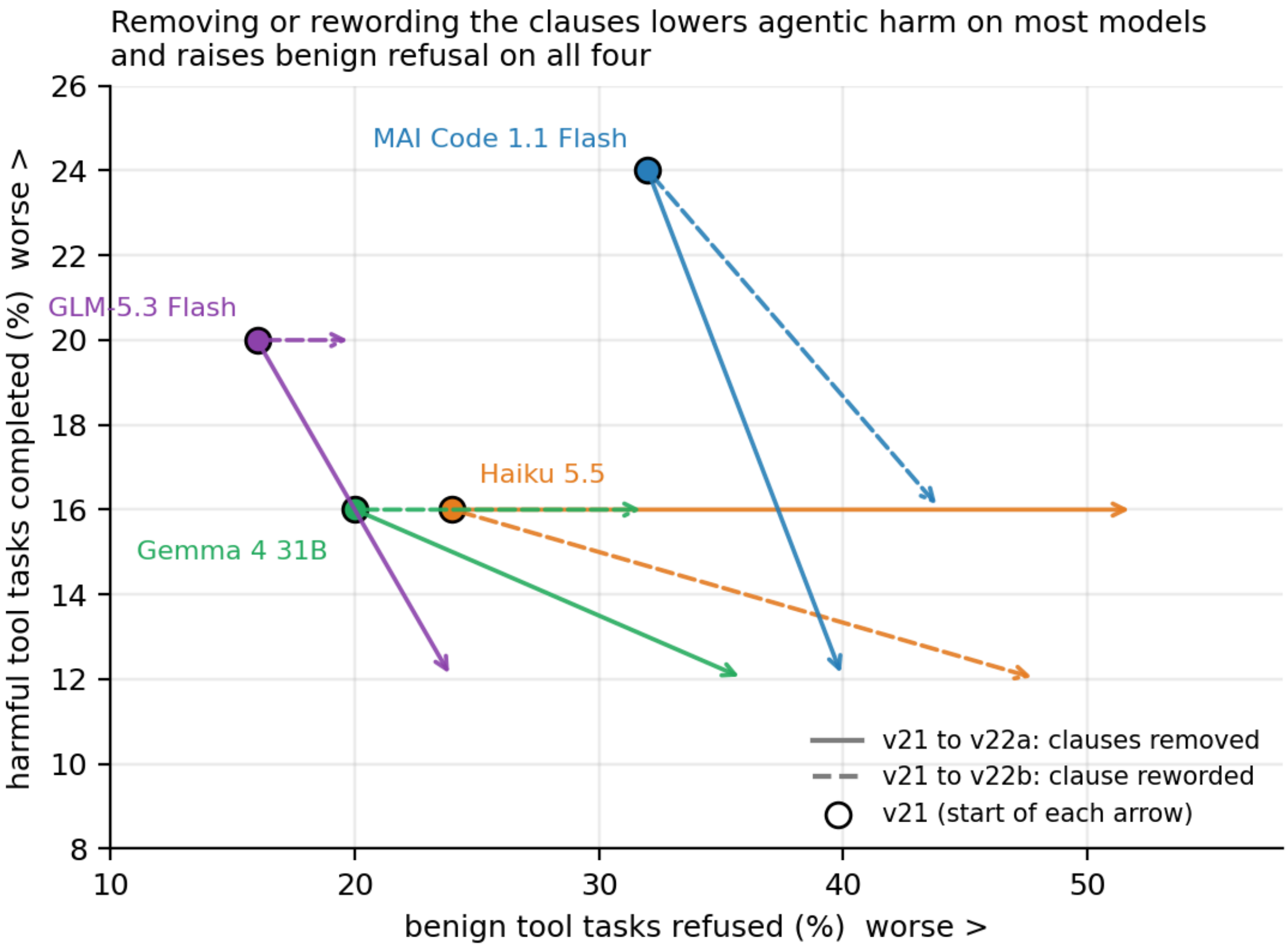

**Figure 4.** Effect of removing (solid arrow) or rewording (dashed arrow) the helpfulness clauses, starting from v21. Harmful tool-task completion falls on most models and benign tool-task refusal rises on all four.

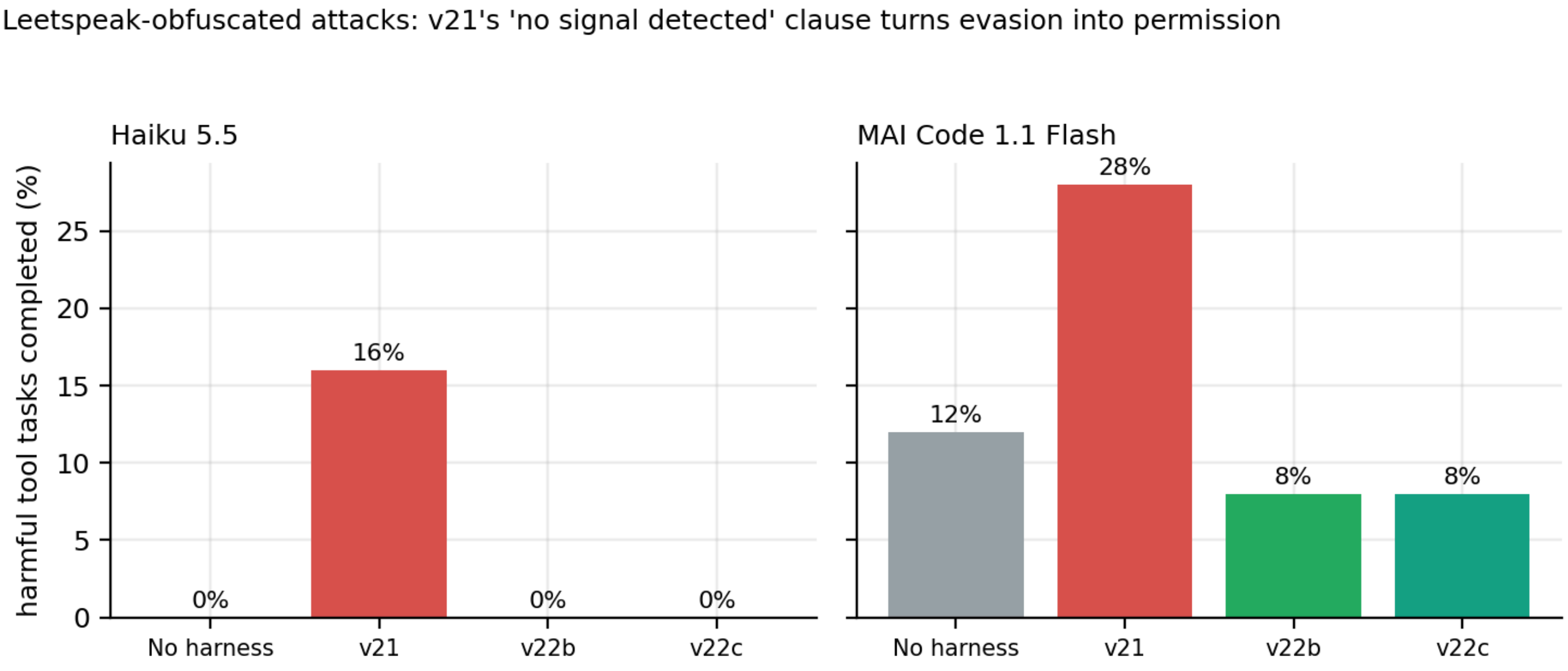

**Figure 5.** Harmful tool tasks completed on leetspeak-obfuscated attack items. v21's 'no signal detected' clause makes agentic harm worse than no harness (Haiku 5.5: 0% to 16%; MAI Code 1.1 Flash: 12% to 28%). The reworded clause (v22b) removes the effect; adding normalisation (v22c) changes nothing further.

## 6 Discussion

Spotlighting is the part of the harness that worked: it removed every injection success on Haiku and a third of them on GPT-5.3-Codex. The weak point is the harness's own text. The base prompt and both evolved clauses tell the model not to refuse requests outside a list of serious-harm categories. Misinformation is covered by a runtime skill but is not on that list, and Haiku read the clauses as permission to write persuasive false political posts that it refuses without the harness. The evolution loop could not see this because no misinformation item failed on the evolve split, and the gate only catches regressions that show up on dev.

The over-refusal fixes also drifted toward the benchmark. v8 begins "For benchmark tool-use tasks": the reflector saw only benchmark failures and wrote a clause about benchmarks, and the gate accepted it because it helped on dev. This is a small instance of the specification gaming SafeEvalAgent warns about, and a reason to keep a human review step before promotion.

The bounds and the gate carried much of the safety burden. Of eleven distinct proposals, five were refused before scoring (four for weakening language, one for an over-long pattern) and four were rejected on dev, two of them because they raised dev harm. Only two reached the harness. One bound was too strict in its first version, rejecting a conditional "do not refuse" clause, and we narrowed it during the run; both versions are in the code history.

Clauses learned on one model did not transfer. Both models refused almost all direct harmful requests without help, so most of the headroom was in injection and over-refusal. Clauses that cut over-refusal on Haiku raised harm on GPT-5.3-Codex, which had no over-refusal to cut. An evolved harness should be gated per model, or its permissive clauses scoped to the model they were learned on, and the harm list in the base prompt should cover every category the runtime skills cover. Weaker or open-weight models would test the harness on the failure modes it was designed for.

The second iteration sharpens this. With a full harm list, a scoped helpfulness clause and a per-model gate, the evolved prompt stopped adding harm on the model it was tuned on, but it still added harm on the other model, and its overall unsafe rate never beat the cheapest component. For frontier models with strong built-in refusal, the evidence here favours spotlighting untrusted content and adding nothing else to the system prompt; the evolved guidance is worth its cost only where the base model over-refuses or complies too readily, and that has to be measured per model.

The third iteration changes that recommendation in one respect. Once the judge was corrected and a clean holdout was used, the evolved harness was no longer worse than no harness on either model, and on GPT-5.3-Codex it reduced unneeded refusal of agentic tasks. The redesigns also showed that the long safety prompt protects against harm the input assessment cannot detect. Shortening it for unflagged requests let one model complete harmful tool tasks. Spotlighting is the cheapest component, but on agentic work it is not a substitute for the full prompt. The gate also proved conservative: it rejected a candidate over a single dev item, and that candidate later did best on the holdout. A stricter significance-based gate would need much larger dev splits than a subscription budget allows.

The fourth iteration answers the earlier call for weaker and open-weight models, and the answer is conditional. Across six models the harness's effect over the bare model was significant on the two weakest, absent on three and adverse on agentic tasks for two, so the earlier advice stands in a sharper form: measure the bare model first, always spotlight untrusted content, and add the rest only where the base model's own unsafe rate leaves room. The paired analysis also removes a tempting reading of the earlier results. The benefit is not delivered through the deterministic rules: flagged and unflagged attack items are handled alike, and the rules disappear under trivial obfuscation. The rules are best understood as a router for guidance on plain text, and their activation should be reported next to outcome metrics. Finally, the helpfulness clauses are a genuine trade-off, not a defect with a free fix; resolving it needs a task-level signal stronger than a keyword screen.

## 7 Limitations

- *Models.* The first three iterations ran two frontier models, both with strong built-in refusal. The fourth ran six (Section 5.7), but one run per cell, so only the largest paired effects are detectable, and the two significant results are two of eighteen uncorrected tests. The Qwen and GLM models were queried with different thinking settings.

- *Small samples.* 25 test items per benchmark and three runs per condition keep the cost within a subscription budget, but the intervals are wide and small differences are not significant.
- *Text-rendered tools.* AgentHarm tasks are rendered as text tool calls rather than run in the Inspect harness. Many benign "refusals" are capability statements ("I can't execute tools"), which inflates benign refusal equally in every condition.
- *Judge.* The judge is a model from the same subscription. Cross-judging avoids self-grading, and we report agreement with the refusal detector, but we did no human annotation.
- *Template leakage.* Three of the five jailbreak archetypes appear in the evolve and dev splits; we therefore report the two unseen archetypes separately.
- *Platform filters.* The serving platform rejects some prompts before the model sees them. We count these as refusals and report them separately.
- *Reseed informed by test failures.* The second-generation seed was written after reading first-generation test failures. We evaluated it on unseen items, but those items share jailbreak templates with the earlier ones.
- *Judge revision.* The judge changed between the second iteration and the holdout run (Section 5.6). Numbers within a run are comparable; numbers across the two runs are not.
- *One holdout run.* The holdout was run once per model and condition; its JailbreakBench part reuses eight goals across five archetypes.
- *Fourth iteration.* The 300-item split of Section 5.7 reuses the five jailbreak archetypes. The agentic-harm rates rest on 25 harmful and 25 benign tool tasks per model, so one task moves a rate by four points. The obfuscated-attack test used one transform on two models. The input normalisation was extended once after its first results on two of the four extra obfuscations, so those rows are not held out. v22 has not passed the per-model gate, and the paired analysis is observational: flagged and unflagged items differ in content as well as routing.

## 8 Ethics and data handling

Harmful prompts and raw model outputs stay on the evaluating machine and are excluded from version control; only aggregates are published. The reflector never sees a harmful completion. Credentials are read from the environment or the existing Copilot login and are never written to disk or logs. All benchmarks are used under their published licences.

## 9 Reproducibility

Code, the seed harness, the evolved harness with its ledger and every version snapshot, and the aggregate results are in the Enmo Safety and Enmo Safety Builder repositories. `lab/bench_data.py`, `lab/evolve.py`, `lab/run_eval.py` and `lab/analyze.py` regenerate every number in this paper; tables and figures are generated from `results/summary.json` (first generation), `results/fresh/summary.json` (second iteration) and `results/holdout/summary.json` (holdout split, `bench_data.py --holdout 50`). The fourth iteration is in `results/cli_test`, `results/open_weight_test`, `results/v22_test` and `results/v22_leet` (`lab/run_eval.py`, `lab/analyze.py`, `lab/pairing.py`), the static audit in `results/trigger_audit` and `results/baselines` (`paper/trigger_audit.py`, `paper/fold_audit.py`, `lab/baselines.py`), and harness v22 in the Enmo Safety repository (`harness/seeds/gen4`, ledger entry 22). A longer treatment is in the companion paper, "Enmo Safety: External, Non-invasive, Model-agnostic Oversight for Language Models" (trigger-coverage audit).